\documentclass[preprint]{aastex}
\usepackage{emulateapj5}
\usepackage{lambdabar}
\usepackage{amsmath}
\usepackage{amssymb}

\newcommand{\sci}[2]{\mbox{$#1 \times 10^{#2}$}}

\newcommand{\lnlambda}{\ln \Lambda}
\newcommand{\units}[1]{\: \mbox{#1}}

\newcommand{\C}{C}
\newcommand{\R}{R}
\newcommand{\NR}{N\!R}
\newcommand{\PFF}{P\!F\!F}

\newcommand{\mycolhead}[1]{\multicolumn{1}{c}{#1}}

\newcommand{\sx}{\hspace{1ex}}

\shorttitle{Gamma-ray emission}
\shortauthors{Hibschman}

\begin{document}
\title{Gamma-ray emission from rotation-powered pulsars}
\author{Johann A.\ Hibschman}
\affil{Theoretical Astrophysics Center, Astronomy Department, University of Califonia, Berkeley}
\affil{601 Campbell Hall, Berkeley, CA 94720-3411}
\email{jhibschman@astron.berkeley.edu}

\begin{abstract}
   Using a simplified model of cascade pair creation over pulsar
   polar caps presented in two previous papers, we investigate the
   expected gamma-ray output from pulsars' low altitude particle
   acceleration and pair creation regions.
   We divide pulsars into several categories, based on which mechanism
   truncates the particle acceleration off the polar cap, and give
   estimates for the expected luminosity of each category.
   
   We find that inverse Compton scattering above the pulsar polar cap
   provides the primary gamma rays which initiate the pair cascades in most
   pulsars.  This
   reduces the expected $\gamma$-ray luminosity below
   previous estimates which assumed curvature gamma ray emission was the
   dominant initiator of pair creation in all pulsars.  Even for
   the brightest pulsars where curvature radiation sets the height of
   the pair formation front (PFF), we
   find predicted luminosities too low to explain the EGRET pulsars,
   suggesting that the source of that emission is an outer magnetosphere
   accelerator.  The predicted polar cap luminosities are large enough,
   however, to be observable by upcoming $\gamma$-ray instruments, which
   provides a firm test for this theory.
\end{abstract}


\maketitle

\section{Introduction}

Although only a few pulsars have so far been detected in $\gamma$-rays,
mostly via the EGRET telescope on CGRO \citep{ulmer94,thompson97},
this number should steadily
grow as a new generation of $\gamma$-ray instruments, such as INTEGRAL
and GLAST, is brought online.  This should greatly assist theoretical
understanding of pulsars, since the highest energy photons
provide a direct window into the underlying mechanisms thought to lead
to pulsar emission of all types.

The primary photon emission mechanisms important in radio pulsars are
inverse Compton scattering (ICS), in both the resonant (RICS) and
Klein-Nishina nonresonant (NRICS) modes, and curvature emission.  The
relative importance of these mechanisms is still unclear.  Most $\gamma$-ray
pulsar papers have concentrated on the curvature emission
\citep{zhang00,romani95,daugherty82}, while an increasing number of
polar-cap physics papers have emphasized the importance of ICS
\citep{dermer90,sturner95,sturner95b,luo96,harding98}.

In the polar-cap acceleration model, particles are extracted from the
polar cap and accelerated by large rotation-induced electric fields,
forming the primary beam.  These particles then emit primary $\gamma$-ray
photons due to ICS and curvature emission, and these photons interact
with the pulsar magnetic field to create electron-positron pairs.  The
density of these secondary pairs increases with height as more and
more photons pair-produce, until the pair density is sufficient to
short out the accelerating field.

Historically, this shorting of the accelerating field was thought to
occur in a thin layer after the creation of the first pair; hence, the region
of no pair creation, $E_\parallel \neq 0$, was thought to be separated
from the region of copious pair creation, $E_\parallel \approx 0$, by
a thin ``pair formation front'' (PFF).  ICS photons, however, create
small numbers of pairs at low altitudes, breaking the connection
between the altitude of the first created pair and the altitude at
which the electric field disappears.  Since the altitude of the first
created pair has no inherent dynamical significance, we use the term PFF in
this paper always as applying to the altitude at which the 
accelerating field is shorted.  Other effects, such as the polarization of
the generated pairs, will also begin at the point of first pair creation, but
these effects must likewise reach some threshold before they affect the
dynamics of the beam.

Once the accelerating field has been shorted out, the primary beam coasts,
continuing to emit $\gamma$-rays.  The total $\gamma$-ray output of a polar
cap is then a combination of the synchrotron $\gamma$-rays produced by
the created secondary particles, the ICS radiation emitted by the
secondary particles, and the primary $\gamma$-rays emitted by the primary
beam.  If the primary beam is radiation-reaction limited, this
emission efficiently converts the beam energy into
$\gamma$-rays; otherwise, only a small fraction is extracted.

We find that the $\gamma$-ray output of pulsars falls into two
categories.  For the majority of pulsars, non-resonant ICS stops the
beam acceleration at small Lorentz factors where $\gamma$-ray emission
is inefficient, leading to low luminosities.  For the remaining
pulsars, the beam is accelerated to high Lorentz factors, resulting in
efficient $\gamma$-ray emission, and a high luminosity.

Using the results of \citet{hibschman01} and \citet{hibschman01b},
henceforth Papers I and II, we
examine the boundary between these categories of pulsars, and predict
the luminosities and spectral characteristics of these objects.

\section{Model}

Inverse Compton scattering depends strongly on the temperature of the
neutron star polar cap, making the thermal cooling model chosen for
the neutron star an important part of the theory.  For simplicity, we
will assume the temperature of the polar cap is entirely due to the
polar cap heating discussed in Paper I.  We use the acceleration model
of \citet{muslimov92,muslimov97}, in the simplified form described in
Paper I, and neglect spatial variations across the polar cap.

This accelerating potential, expressed in units of $mc^2/e$ so as to
give the expected particle Lorentz factor, is
\begin{equation}
  \begin{array}{rclll}
    \Phi_{low}(t)  & = & \Phi_1 t^2 & mc^2 e^{-1}, & t < 1 \\
    \Phi_{high}(t) & = & \Phi_1 t   & mc^2 e^{-1}, & t > 1
  \end{array}
\end{equation}
where $\Phi_1 = \sci{5.14}{4}\,B_{12} P^{-5/2}$, $t \equiv s/s_1$, and
$s_1 = \sci{8.87}{-3} P^{-1/2} R_*$, while $s$ is the altitude above
the stellar surface.

Although we use general relativity though the Muslimov and Tsygan 
accelerating potential, we neglect other relativistic effects, such as 
the changes to the magnetic field near the surface and the gravitational 
red-shift of the emitted photons. These effects are of order 10--15\%, 
so including them would pretend to greater accuracy than justified. The 
accelerating potential is unique, as it is only the charge difference 
created by the relativistic contribution which creates the starvation 
electric field.

\subsection{Emission rates}

First, we consider the emission of a single primary beam particle,
moving along the field lines above the pulsar cap.  We neglect, for
now, discussion of the secondary particles created by pair-production,
as the total energy emitted by these particles is clearly limited by
that emitted by the primaries.  For this section, we effectively
assume that only a negligible fraction of the energy emitted in
primary $\gamma$-rays remains in the generated electron-positron
plasma.  From the results of Paper II, this naturally follows in
low-$B$ pulsars ($B_{12}$ < 1) and results in other pulsars
due to the RICS of the secondary pair plasma.

Since NRICS is only logarithmically-dependent on the Lorentz factor of
the beam, the NRICS power emitted is limited primarily by the
attenuation of the background thermal photons as the beam particles
move away from the star.  If we assume a hot polar cap of angular
radius $\theta_c = \sqrt{\Omega R_*/c}$, where $\Omega$ is the angular
velocity of the pulsar, $\Omega = 2\pi/P$ and $R_*$ is the radius of
the neutron star, assumed to be 10 km, the total energy radiated by one
particle is
\begin{equation}
  E_{\NR} = \frac{\theta_c R_*}{c} P_{\NR} = \sci{6.1}{-4} \, P^{-1/2}
    T_6^2 \units{ergs}
  \label{eq:E_nr}
\end{equation}
The approximate form used for the NRICS power, $P_{\NR}$, is that
given in Paper I, and $T_6$ is the temperature of the polar cap,
in units of $10^6$ K.

The power emitted via RICS, while only logarithmically sensitive to
the decline in the thermal photon flux, decreases quickly with
increasing Lorentz factor.  Because of this dependence, most of the
RICS power is emitted at low altitudes where the Lorentz factor is
still small.  The expected total output per particle is
\begin{equation}
  E_{\R} = \frac{s_{min}}{c} P_{\R}(\gamma_{min}) = \sci{4.5}{-5} \,
    P^{3/4} B_{12} T_6^{3/2} \units{ergs}
  \label{eq:E_r}
\end{equation}
where $\gamma_{min} = \epsilon_B / kT$ is the minimum Lorentz factor
at which thermal photons are upscattered into resonance with the beam,
$s_{min}$ is the altitude at which the beam particles reach
$\gamma_{min}$, and $P_{\R}$ is the power emitted by RICS emission,
as given in Paper I.  In the acceleration model from Paper I, $s_{min} =
450 \, P^{3/4} T_6^{-1/2}$ meters, assuming a star of radius 10 km.
 
The power emitted by curvature radiation is strongly dependent on the
Lorentz factor of the beam, varying as $\gamma^4$, so most of the
energy emitted by curvature radiation is emitted as the beam coasts
above the PFF.  Once particle acceleration stops, the Lorentz factor
of the primary beam declines according to
\begin{equation}
  \gamma(s > s_{\PFF}) = \gamma_{\PFF} \left(1 + 3
    \frac{P_{\C}(\gamma_{\PFF})}{\gamma_{\PFF}} \frac{R_*}{c}
    \ln \frac{1+s}{1+s_{\PFF}}
    \right)^{-1/3},
  \label{eq:gamma_s}
\end{equation}
assuming a dipolar magnetic field. The upper limit may be estimated by
finding the altitude at which the high-energy primary beam decouples
from the magnetic field.  Equating the energy density in the beam to
the energy density in the magnetic field gives a decoupling height of
$r_{max} = 4123 B_{12}^{1/3} P^{1/3} \gamma_7^{-1/3} R_*$, where
$\gamma_7$ is the Lorentz factor of the beam in units of $10^7$.  The
PFF is close to the surface, so the logarithm above is approximately
8.3.

The total curvature energy emitted is then
\begin{equation}
  E_{\C} = (\gamma_{\PFF} - \gamma(s_{max})) mc^2 = 8.2
    (\gamma_{\PFF7} - \gamma_{7}(s_{max})) \units{ergs}.
  \label{eq:E_c}
\end{equation}

Comparing the emitted energies, equations (\ref{eq:E_nr}),
(\ref{eq:E_r}), and (\ref{eq:E_c}), we find that for typical Lorentz
factors of $\gamma_{\PFF} > 10^5$, only the curvature emission may
radiate any appreciable fraction of the beam particle energy.  The
minimum Lorentz factor at which radiation reaction is important is
then
\begin{equation}
  \gamma_{RR} = \sci{2.33}{7}\, P^{1/3},
  \label{eq:g_rr}
\end{equation}
which is the Lorentz factor at which the primary beam particles lose
half their energy to curvature radiation.  Above this Lorentz factor,
roughly all of the energy in the beam is lost to $\gamma$-rays; below,
the beam propagates without significant radiation losses.

The expected curvature energy loss in these two regimes is
\begin{equation}
  E_{\C} \approx \gamma_{\PFF} \: mc^2 = \sci{8.2}{-7} \gamma_{\PFF}
  \units{ergs}
  \label{eq:E_c_high}
\end{equation}
if radiation reaction is important and
\begin{equation}
  E_{\C} \approx \frac{8.3 R_*}{c} P_{\C}(\gamma_{\PFF}) = 
   % \sci{1.84}{-22} \, P^{-1} \gamma_{\PFF}^4
   \sci{1.51}{-28} \, P^{-1} \gamma_{\PFF}^4 \units{ergs}
  \label{eq:E_c_low}
\end{equation}
if not.

\subsection{Luminosity}

Using the polar cap model from Paper I, we can classify pulsars 
according to the
mechanism which sets the PFF and whether the beam is radiation
reaction limited ($\gamma_{\PFF} > \gamma_{RR}$).  The pair formation
model then gives the altitude of the PFF, $s_{\PFF}$, and the Lorentz
factor at that altitude, $\gamma_{\PFF}$.

Given $\gamma_{\PFF}$, we can compute the total expected luminosity by
multiplying the total energy emitted by a single beam particle by the
number of particles emitted by the polar cap, $\dot{N} = n_{GJ} c
\theta_c R_*^2 = \sci{1.37}{30} B_{12} P^{-2}\,s^{-1}$, where $n_{GJ}$
is the expected Goldreich-Julian number density, $n_{GJ} = \Omega
\cdot B / 2 \pi c e$.

As a slight modification to the model in Paper I, we find that,
in comparison with the numerical results of Paper II, the PFF
from curvature emission is more accurately found by finding the
altitude at which the first pair is formed.  This is due to the
steadily increasing intensity of curvature emission with increasing
Lorentz factor.

A curvature photon emitted at altitude $s$ will pair produce at
\begin{equation}
  s_{\pm} = s + \frac{1}{4} \frac{\epsilon_a}{\epsilon_{\C}(\gamma(s))} R_*,
\end{equation}
where $\epsilon_{\C}$ is the typical curvature photon energy,
$\epsilon_{\C} = \sci{5.8}{-19} \gamma^3 \rho_8^{-1} mc^2$, and
$\epsilon_a$ is the scaling energy for pair production from Paper I,
$\epsilon_a = 2166 \, B_{12}^{-1} P^{1/2} f_\rho$. Here $f_\rho$
is the ratio of the actual field line radius of curvature to the
radius of curvature of the dipole field line which intersects the
stellar surface at $\theta_c$; for the remainder of the paper, this is
taken to be 1.

The minimum value of this is at
\begin{equation}
  s_{\PFF,\C} = 1.91 \, B_{12}^{-1} P^{7/4} f_\rho^{1/2} R_*,
  \label{eq:s_pff_c}
\end{equation}
if the PFF takes place in the linear regime of the accelerating
potential at $s \gtrsim \theta_c R_*$, or
\begin{equation}
  s_{\PFF,\C} = 0.211 \, B_{12}^{-4/7} P^{11/14} f_\rho^{2/7} R_*,
  \label{eq:s_pff_c_low}
\end{equation}
if in the quadratic regime at $s \lesssim \theta_c R_*$.  

Using the semi-numerical model of Paper II, we then classify pulsars
by which the emission mechanism produced the PFF, by whether the PFF
occurred at low or high altitude (in comparison with the polar cap
width), and by whether the beam was radiation-reaction limited.  In
principle, this gives 12 categories; in practice, there are only five
important divisions.

Out of the 540 pulsars in the Princeton pulsar catalog with positive
$\dot{P}$, the majority, 315, had a PFF set by NRICS at high
altitude with $\gamma_{\PFF} < \gamma_{RR}$.  For all of these
objects, curvature radiation is the primary energy loss mechanism in the
beam drift region above the PFF.  Although most of the radiated energy
is curvature emission, it is the relatively sparse, but individually
much higher-energy, NRICS photons that first pair produce and set
the PFF.

In the model of Paper I, self-consistent cap heating produces a PFF
height and final Lorentz factor of
\begin{eqnarray}
  s_{\PFF,\NR,pc}      & = & 0.749 \, B_{12}^{-4/3} P^{11/6} R_* \\
  \gamma_{\PFF,\NR,pc} & = & \sci{4.34}{6} \, B_{12}^{-1/3} P^{-1/6}
\end{eqnarray}
which corresponds, through equation
(\ref{eq:E_c_low}) to a total luminosity of
\begin{equation}
  L_{\NR} = \sci{7.3}{28} \, B_{12}^{-1/3} P^{-11/3}
  \units{erg s$^{-1}$}.
  \label{eq:l_nr}
\end{equation}

This is much lower than previous estimates of pulsar $\gamma$-ray
luminosity, and is by far the most common case.  However, this
estimate is only strictly accurate for magnetic fields less than or on
the order of $10^{12}$ Gauss due to the increasing fraction of
particle energy left in the particles at higher values of $B$, as
discussed in Paper II.  For higher fields, the luminosity will be
larger, as NRICS is less efficient, but still less than the value
expected from pure curvature radiation.

The next most common case is high-altitude inertially limited RICS,
accounting for 126 of the 540 pulsars.  For most of these pulsars, the
$\gamma$-ray emission in the beam drift region above the PFF
is again dominated by curvature emission; the
23 of these where this is not true are among the least luminous
pulsars, and so we neglect them.

From Paper I, the PFF height and final Lorentz factor in this case are
\begin{eqnarray}
  s_{\PFF,\R,pc} & = & 9.66 \, B_{12}^{-16/7} P^{1/2} R_* \\
  \gamma_{\PFF,\R,pc} & = & \sci{5.60}{7} \, B_{12}^{-9/7} P^{-3/2}
\end{eqnarray}
which corresponds, via equation (\ref{eq:E_c_low}), to a total luminosity
of
\begin{equation}
  L_{\R}^{high} = \sci{2.0}{33} \, B_{12}^{-29/7} P^{-9}
  \units{ergs s$^{-1}$}.
  \label{eq:l_r_high}
\end{equation}

The third category combines those pulsars where either curvature or
NRICS sets the PFF at high altitude and where the beam is radiation
reaction limited, for a total of 57 objects.  In practice, we find
that these pulsars are all well-modeled by the expected PFF for
curvature emission, equation (\ref{eq:s_pff_c}), and equation
(\ref{eq:E_c_high}) for the energy loss.  This gives a final Lorentz
factor and total luminosity of
\begin{eqnarray}
  \gamma_{\PFF,\C}^{high} & = & \sci{1.10}{7} \, P^{-1/4} \\
  L_{\C}^{high} & = & \sci{1.2}{31} \, B_{12} P^{-9/4}
  \units{ergs s$^{-1}$}
  \label{eq:l_c_high}
\end{eqnarray}
which is essentially the same as the \citet{zhang00} value for their
regime II.  As they mentioned, this preserves the empirical $L_\gamma
\propto L_{SD}^{1/2}$ relation.  These are among the brightest
$\gamma$-ray pulsars, behind only category 4 in total luminosity.

The fourth category is similar to the third, in that it includes
radiation reaction limited beams where the PFF is set by both
curvature and NRICS emission, but at low altitudes rather than high.
These are the 21 brightest, highest-potential pulsars.  Using
equations (\ref{eq:s_pff_c_low}) and (\ref{eq:E_c_high}) gives a
Lorentz factor and luminosity of
\begin{eqnarray}
  \gamma_{\PFF,\C}^{low} & = & \sci{2.90}{7} \, B_{12}^{-1/7} P^{1/14} \\
  L_{\C}^{low} & = & \sci{3.3}{31} \, B_{12}^{6/7} P^{-27/14}
  \units{ergs s$^{-1}$}
  \label{eq:l_c_low}
\end{eqnarray}
which corresponds to regime I of \citet{zhang00}.

The final category consists of the 13 pulsars where RICS sets the PFF
in the low-altitude regime, with an inertially limited beam.  These
are all pulsars with fields well in excess of $10^{13}$ Gauss, field
strengths beyond the expected range of validity of this theory.  The
the low-altitude PFF height and Lorentz factor are
\begin{eqnarray}
  s_{\PFF,\R,pc}^{low} & = & 4.03 \, B_{12}^{-2} P^{3/8} R_* \\
  \gamma_{\PFF,\R,pc}^{low} & = & \sci{1.06}{10} \, B_{12}^{-3} P^{-3/4}.
\end{eqnarray}  
For these pulsars, RICS emission produced more gamma-rays than curvature,
so using the emitted energy from equation (\ref{eq:E_r}), we find a
luminosity of
\begin{equation}
  L_{\R}^{low} = \sci{7.2}{26} \, B_{12}^{5/4} P^{-67/32}
  \units{ergs s$^{-1}$}.
  \label{eq:l_r_low}
\end{equation}

As in Paper I, the emission mechanism with the smallest PFF height is
the dominant mechanism.  Owing to the multiplicity of categories,
generalizations are difficult, but we can derive a few formulae for
the boundaries between regimes by considering only the high-altitude
limits.

In this simplest appoximation, the boundary between the NRICS-dominated
and curvature-dominated pulsars is at $B_{12} = 0.061\,P^{1/4}$.  Pulsars
with weaker magnetic fields are dominated by curvature, and equation
(\ref{eq:l_c_high}) is the appropriate luminosity,
while pulsars with fields stronger than this are dominated by NRICS
and equation (\ref{eq:l_nr}) is appropriate.  However, this
ignores the weakening of NRICS with increasing magnetic field discussed
in Paper II, and so should be taken as a statement that the millisecond
pulsars are certainly controlled by curvature, while the higher-field,
longer-period pulsars are favored by NRICS, but must be examined carefully,
using either the full algebraic results from Paper I or the semi-numerical
model of Paper II.

The boundary
between NRICS and RICS lies at $P_{\NR,\R} = 6.81 \, B_{12}^{-5/7}$, with
higher periods favoring RICS, while that between RICS and curvature
is at $P_{\R,\C} = 3.66 \, B_{12}^{-36/35}$, again with RICS dominating
at longer period.  For pulsars with magnetic fields larger than
approximately \sci{4}{12} Gauss, NRICS is ineffective in setting the
PFF, and the only
active mechanisms are RICS and Curvature; at mid-range fields between
\sci{3}{11} Gauss and \sci{4}{12} Gauss, NRICS sets the PFF, while at
lower fields, curvature sets it.

\subsection{Flux}

Since in this model the observed $\gamma$-rays originate from particles
moving along the field lines above the pulsar polar cap, the predicted
luminosities translate directly into a predicted flux.
Since the polar cap is the source of the $\gamma$-rays,
they are beamed into a cone of opening angle $\theta \approx (3/2)
\theta_c$.  In general, this opening angle will vary with the altitude
of the emission, but for most pulsars the altitude of the PFF is small
compared to the stellar radius, so simply evaluating the opening angle
at the surface is sufficient.
If the
luminosity is spread evenly throughout this cone, this produces a peak
flux of
\begin{equation}
  \phi_{peak} = \frac{L}{\pi \theta^2 d^2}
\end{equation}
where $d$ is the distance to the pulsar.  This is the flux seen while
looking ``down the barrel of the gun.''  Only if the spin axis of the
pulsar is nearly aligned with the line of sight will the observed
flux be on this order.  The average flux is reduced by $\theta/\pi$
if the spin axis is perpendicular to the line of sight and by
approximately $\theta / \pi \sin \alpha$ in the general case,
where $\alpha$ is the angle between the magnetic moment and the rotation
axis and we have effectively assumed that the observer's line of sight
passes through the center of the emitting cone ($\beta = 0$).  This
represents the fraction
of the pulse period where the emitting cone is directed towards the
observer.
\begin{equation}
  \phi_{ave} = \frac{L}{2 \pi^2 \theta \sin \alpha d^2}.
\end{equation}

\begin{figure*}
  \plotone{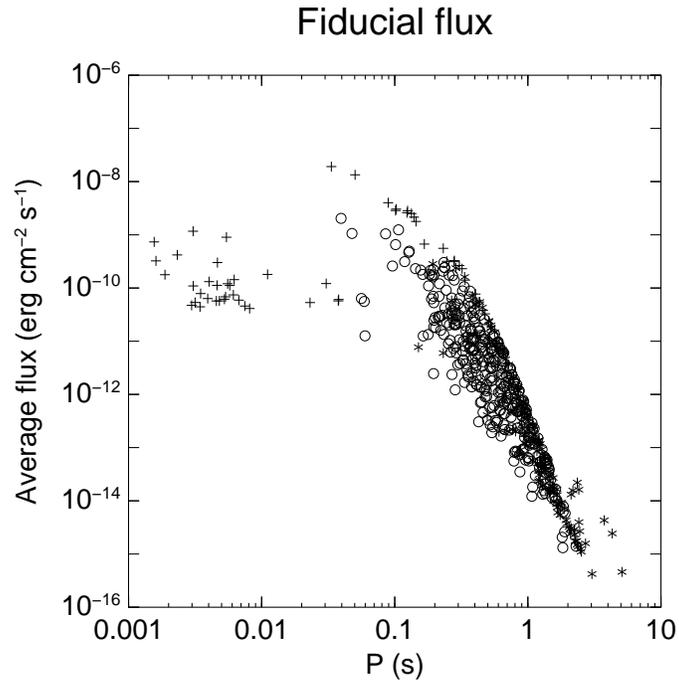}
  \caption{$\gamma$-ray flux expected from pulsars, as a function of
    pulse period, if
    all pulsars are placed at a fiducial distance of 1 kpc.}
  \label{fig:fiducial_flux}
\end{figure*}

\begin{figure*}
  \plotone{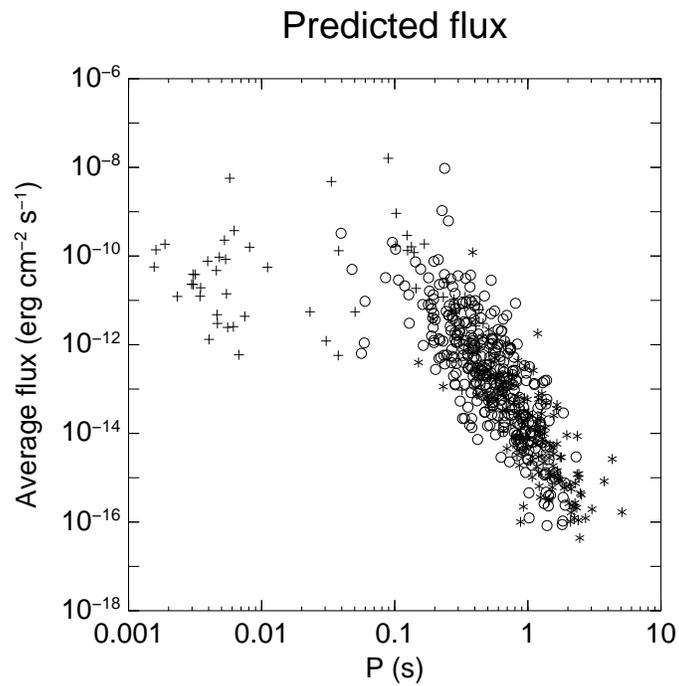}
  \caption{$\gamma$-ray flux expected from pulsars, as a function of
    pulse period, given the actual
    pulsar distances.}
  \label{fig:obs_flux}
\end{figure*}

This average flux is the observable, not the total luminosity.  In
terms of the luminosities from the previous section, this flux is
\begin{equation}
  \phi_{ave} = \sci{3.7}{-43} P^{1/2} (\sin \alpha)^{-1} d_{kpc}^{-2}
    L \units{ergs cm$^{-2}$ s$^{-1}$}.
  \label{eq:lum_to_flux}
\end{equation}
Since the $\sin \alpha$ term is a factor of order unity,
and these approximations are only good to approximately a factor of
2, we simply set it to 1 for the remainder of the paper.
We plot the expected observable flux as a function
of the pulsar period using a fiducial pulsar distance of 1 kpc in
Figure \ref{fig:fiducial_flux}, to show the inherent dependencies on
pulsar parameters, and using the estimated distances of each individual
object
in Figure \ref{fig:obs_flux}, to show the predicted observable flux.

\section{Spectral shape}

In the previous section, we discussed the total energy output of the
pulsar; here, we turn to the expected shape of the spectrum itself.
In Paper II, we found that a cascade of pair creation from a single
absorbed photon produces a response with a power law index of $-3/2$,
due to the reprocessing of synchrotron photons.  In order to be
absorbed, however, the photon must have an energy greater than
$\epsilon_{min} = 5134 \, B_{12}^{-1} P^{1/2} mc^2$, presuming that
the photon was emitted parallel to the magnetic field at the edge
of the polar cap at the surface of the star.

Since most of the power emitted in $\gamma$-rays from pulsar polar
caps is due to curvature emission, except for the few extreme high-field
objects where RICS dominates, the energetics of the curvature
photons determine the shape of the spectrum.  The minimum Lorentz
factor for curvature pair production may be found by equating
$\epsilon_C(\gamma) = \epsilon_{min}$, yielding
\begin{equation}
  \gamma_{\C} = \sci{2.01}{7} \, B_{12}^{-1/3} P^{1/3}
\end{equation}
which is the Lorentz factor at which the critical curvature energy
$\epsilon_{\C}(\gamma) = \sci{5.8}{-19} \rho_8^{-1} \gamma^3 mc^2$ equals
the minimum energy to pair produce, $\epsilon_{min}$, assuming a
dipole field radius of curvature.

If $\gamma_{\PFF} > \gamma_{\C}$, then the copious curvature photons will
pair-produce, and the observed radiation will be the synchrotron
emission of the generated pairs.  At low energies, $\epsilon <
\epsilon_a / (1+a^2) \lnlambda$ in the notation of Paper II, the spectral
index is the $-2/3$ of
unprocessed synchrotron radiation, while at higher energies,
% energies, $\epsilon_a / (1+a^2) \lnlambda < \epsilon < \epsilon_a$,
the spectral index is the characteristic $-3/2$.
%, with an exponential
%tail at higher energies.

If $\gamma_{\PFF} < \gamma_{\C}$, then the curvature photons will be
observed directly.  The beam energy loss in this case, for all pulsars
examined, is small enough that the characteristic curvature energy
remains effectively
unchanged.  Therefore, the unmodified $-2/3$ spectrum of curvature radiation
will be seen, extending from low energies up to
$\epsilon_{\C}(\gamma)$.  This energy will
clearly be lower than the
traditional estimate of the cut-off energy of $\epsilon_a$ and
dependant on the mechanism which sets the PFF.

\begin{figure*}
  \plotone{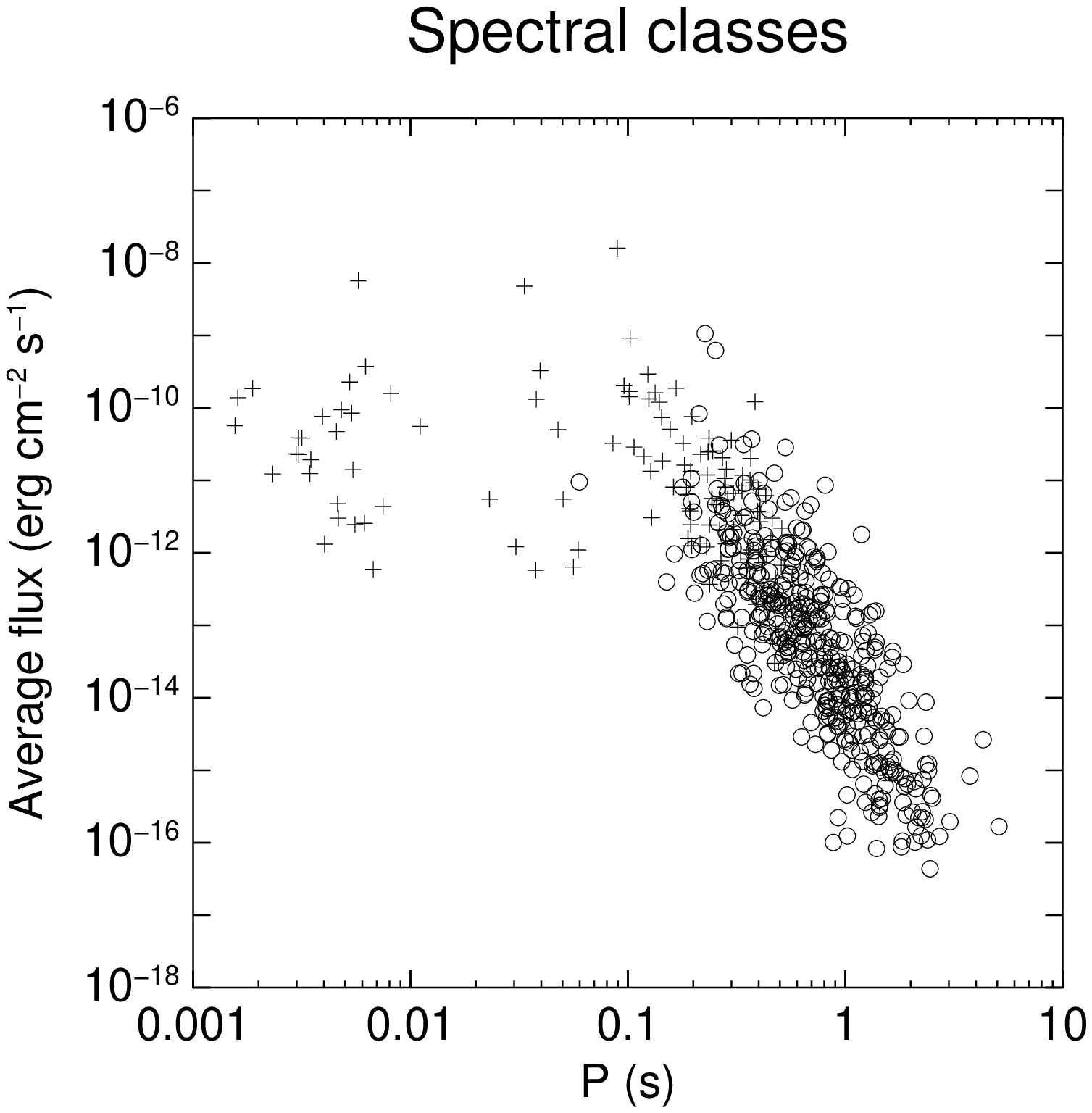}
  \caption{Expected observed $\gamma$-ray flux vs. pulsar period.  Each
    pulsar has been classified into those where curvature radiation is
    expected to produce pairs ({\it crosses}) and those where it is not
    ({\it circles}).  Pulsars where curvature emission produces pairs are
    expected to have a spectral index of $-3/2$, while those which do
    not are expected to have an index of $-2/3$.}
  \label{fig:curv_p}
\end{figure*}

\begin{figure*}
  \plotone{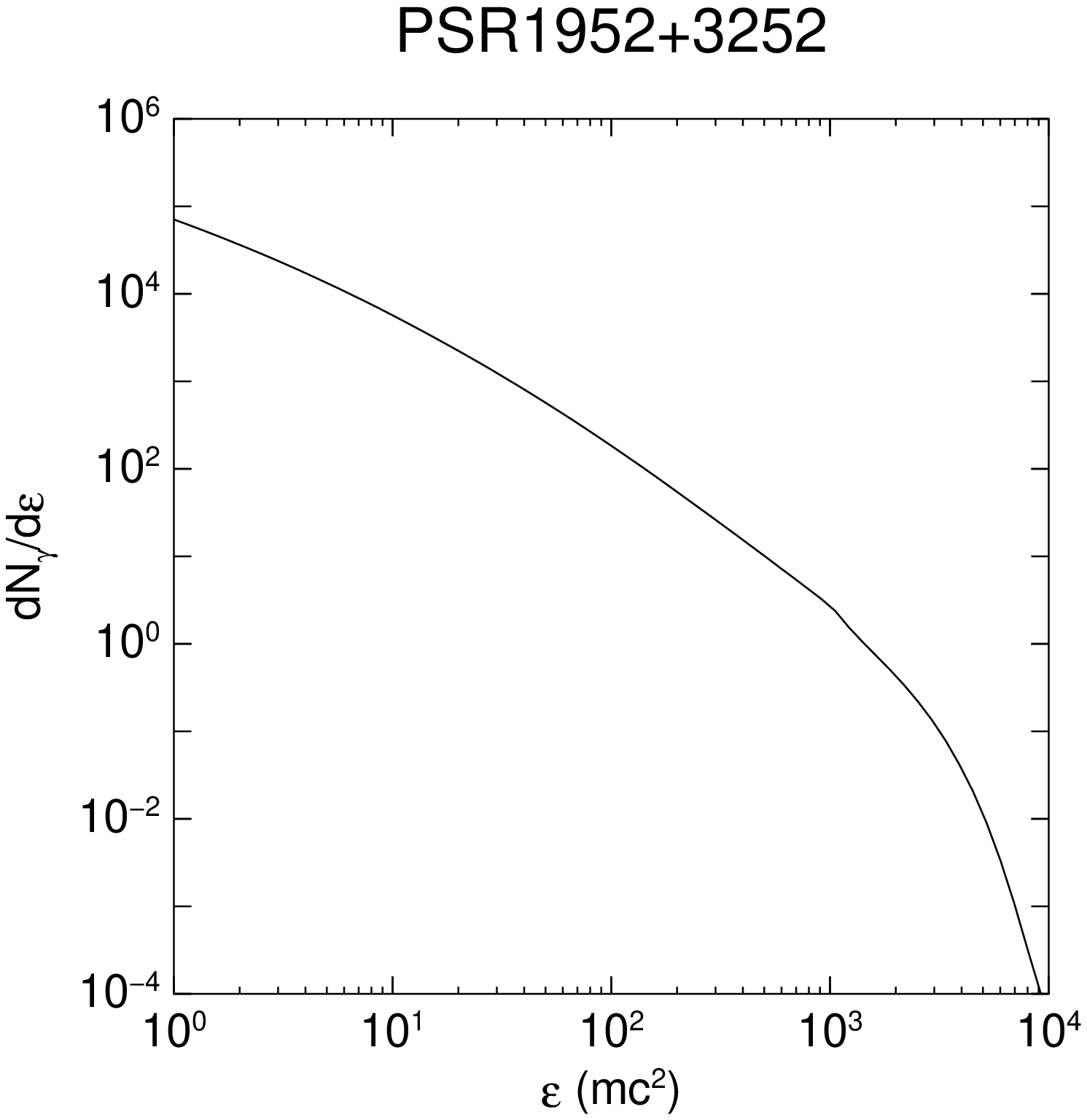}
  \caption{Predicted $\gamma$-ray spectrum produced by a single
    primary particle for 1952+3252, illustrating
    the strong curvature-induced pair production regime, with the resultant
    $\nu = -3/2$ power-law spectrum.  The spectrum is given in units
    of $N/mc^2$.}
  \label{fig:1952-1932}  
\end{figure*}

Figure \ref{fig:curv_p} shows the results of dividing the pulsars into
two categories, based on this division.  Out of the top 20 brightest
expected polar-cap $\gamma$-ray pulsars, only two are expected to have
the $-2/3$ power law, J0953+0755 and J1932+1059.  To illustrate the
difference, we plot in Figure \ref{fig:1952-1932} the
numerically-calculated $\gamma$-ray spectrum of J1952+3252, which is
expected to have a $-3/2$ spectrum, and J1932+1059.  The numerical
method used was that described in Paper II, and the results confirm
the expectations.

\begin{figure*}
  \plotone{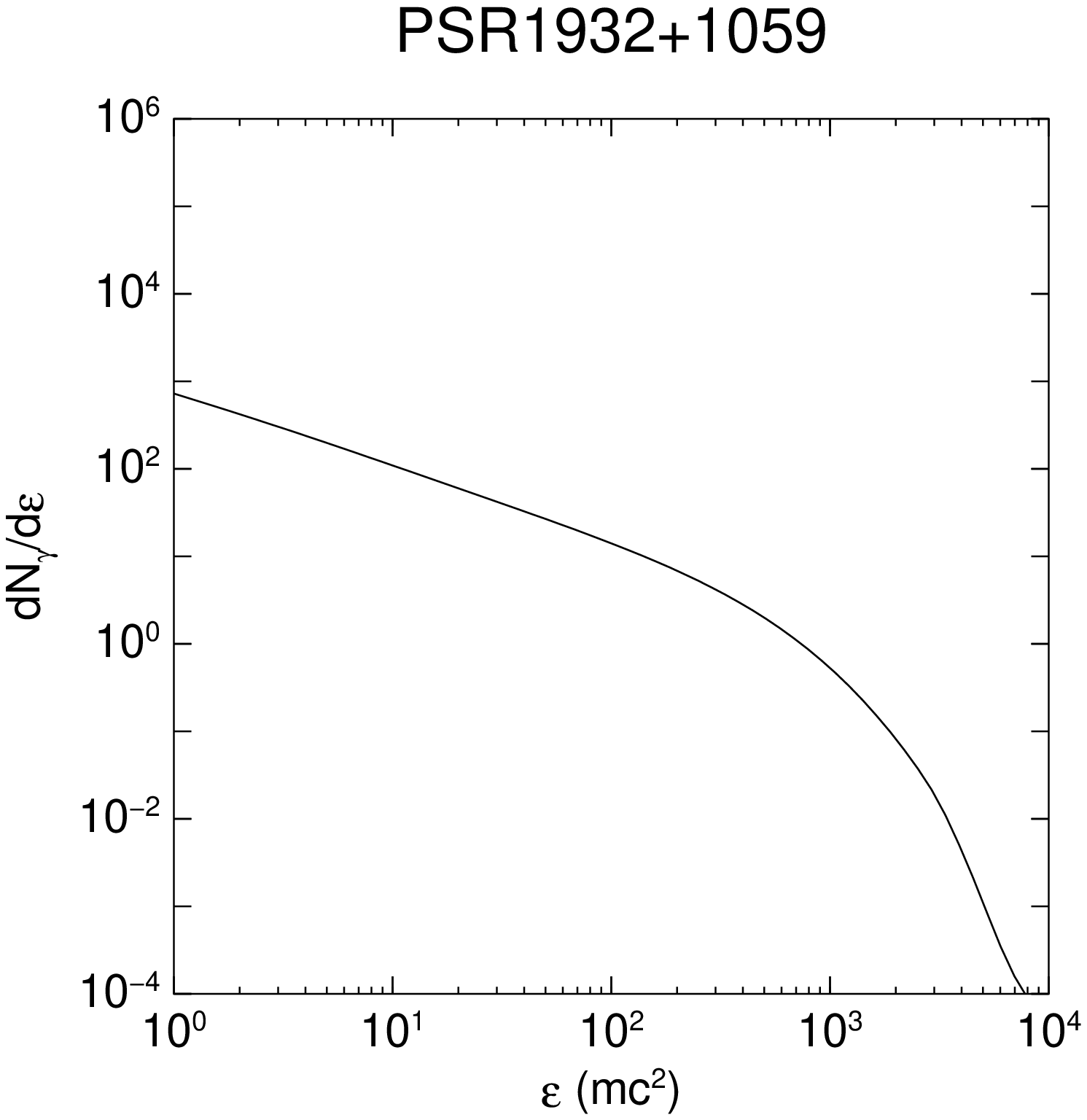}
  \caption{Predicted $\gamma$-ray spectrum produced by a single
    primary particle for 1932+1059, illustrating
    the sparse pair production regime, with the resultant
    $\nu = -2/3$ power-law.  The spectrum is given in units
    of $N/mc^2$.}
\end{figure*}

The maximum energies of these spectra depend on whether the beam
is radiation-reaction limited or not, not on the minimum energy for
pair production, $\epsilon_a$.  All photon energies
higher than $\epsilon_a$ are converted by pair production to photons of
lower energy, but $\epsilon_a$ steadily increases with altitude as
$r^3$, eventually surpassing the maximum energy of the raw photon spectrum
emitted by the beam particles.  At that point, and beyond, the maximum
energy of the spectrum is set by the energy of the beam itself.

Since curvature radiation is the strongest emission mechanism for
all of the brightest pulsars and all but a small minority of the others,
the maximum energy will be the characteristic curvature energy, evaluated
at the high-altitude coasting Lorentz factor of the beam.  However, the
relevant Lorentz factor is slightly different from that calculated in
the previous section, equation (\ref{eq:g_rr}). Since the radius of
curvature of the field steadily increases with altitude, the curvature
energy of a coasting beam will steadily decrease, as $r^{-1/2}$.  The
maximum energy observed will then arise from emission from an intermediate
regime; high enough that the magnetic field no longer absorbs photons,
but low enough that the radius of curvature is still small.  In this case,
the appropriate limiting Lorentz factor is where the scale height for energy
loss is equal to the stellar radius, $\gamma'_{RR} = \sci{3.56}{7} \,
P^{1/3}$.

If the beam is radiation-reaction limited in this sense, i.e.
if $\gamma_{\PFF} > \gamma'_{RR}$, then the maximum energy is the curvature
energy evaluated at $\gamma'_{RR}$, or
\begin{equation}
  \epsilon_{max}^{RR} = \epsilon_{\C}(\gamma'_{RR}) = 10.2 \, P^{1/2} \units{GeV}
\end{equation}
where we have evaluated the curvature energy at a radius of $2 R_{*}$.

If the beam is not radiation-reaction limited, then the maximum energy
depends on the Lorentz factor of the beam, through
\begin{equation}
  \epsilon_{max}^{coast} = \epsilon_{\C}(\gamma_{\PFF}) =
    \sci{3.2}{-22} \, P^{-1/2} \gamma_{\PFF}^3 \units{GeV}.
\end{equation}

The expected observable fluxes and maximum energies for the 25 brightest
pulsars are shown in Table \ref{table:brightest}.

%\begin{deluxetable}{lrlccccl}
%   \tablecaption{Brightest pulsars}
%   \tablehead{  \colhead{} & \colhead{$P$} & \colhead{$B$} & \colhead{PFF} &
%     \colhead{RR}  & \colhead{Curv}  & \colhead{$\epsilon_{max}$} &
%     \colhead{Flux} \\
%     \colhead{Name} & \colhead{(s)} & \colhead{(Gauss)} & \colhead{mech.} &
%     \colhead{limited?} & \colhead{pairs?} & \colhead{(GeV)} &
%     \colhead{(erg cm$^{-2}$ s$^{-1}$)}}
%   \tablewidth{0pc}
%   \startdata

\begin{table*}
  \caption{Brightest pulsars \label{table:brightest}}
  \begin{center}
  \begin{tabular}{lrcccccc}  \tableline \tableline
     \mycolhead{} & \mycolhead{$P$} & \mycolhead{$B$} & \mycolhead{PFF} &
     \mycolhead{RR}  & \mycolhead{Curv.}  & \mycolhead{$\epsilon_{max}$} &
     \mycolhead{Predicted Flux} \\
     \mycolhead{Pulsar} & \mycolhead{(s)} & \mycolhead{(Gauss)} & \mycolhead{mech.} &
     \mycolhead{limited?} & \mycolhead{pairs?} & \mycolhead{(GeV)} &
     \mycolhead{(erg cm$^{-2}$ s$^{-1}$)} \\ \tableline
  J0835$-$4510 \rule[0mm]{0mm}{2.5ex} &  0.089 & \sci{6.8}{12} & Curv  & Yes & Yes &  3.07 & \sci{1.6}{-8\sx} \\ 
  J0633$+$1746 & 0.237 & \sci{3.3}{12} & NRICS & Yes & Yes &  3.73 & \sci{9.5}{-9\sx} \\ 
  J0437$-$4715 & 0.006 & \sci{1.2}{9\sx} & Curv  & Yes & Yes &  0.78 & \sci{5.7}{-9\sx} \\ 
  J0534$+$2200 & 0.033 & \sci{7.6}{12} & Curv  & Yes & Yes &  1.88 & \sci{4.8}{ -9\sx} \\ 
  J1932$+$1059 & 0.227 & \sci{1.0}{12} & NRICS & No  & No  &  0.81 & \sci{1.1}{-9\sx} \\ 
  J1709$-$4428 & 0.102 & \sci{6.3}{12} & Curv  & Yes & Yes &  3.29 & \sci{9.2}{-10} \\ 
  J0953$+$0755 & 0.253 & \sci{4.9}{11} & NRICS & No  & No  &  0.55 & \sci{6.2}{-10} \\ 
  J1300$+$1240 & 0.006 & \sci{1.7}{9\sx} & Curv  & Yes & Yes &  0.81 & \sci{3.7}{-10} \\ 
  J1952$+$3252 & 0.040 & \sci{9.7}{11} & NRICS & Yes & Yes &  2.04 & \sci{3.3}{-10} \\ 
  J1048$-$5832 & 0.124 & \sci{7.0}{12} & Curv  & Yes & Yes &  3.61 & \sci{2.9}{-10} \\ 
  J1012$+$5307 & 0.005 & \sci{5.6}{8\sx} & Curv  & Yes & Yes &  0.74 & \sci{2.3}{-10} \\ 
  J2043$+$2740 & 0.096 & \sci{7.0}{11} & NRICS & Yes & Yes &  3.18 & \sci{2.0}{-10} \\ 
  J0742$-$2822 & 0.167 & \sci{3.4}{12} & Curv  & Yes & Yes &  4.19 & \sci{1.9}{-10} \\ 
  J0034$-$0534 & 0.002 & \sci{2.3}{8\sx} & Curv  & Yes & Yes &  0.44 & \sci{1.9}{-10} \\ 
  J1826$-$1334 & 0.101 & \sci{5.6}{12} & Curv  & Yes & Yes &  3.27 & \sci{1.7}{-10} \\ 
  J1803$-$2137 & 0.134 & \sci{8.6}{12} & Curv  & Yes & Yes &  3.75 & \sci{1.6}{-10} \\ 
  J1730$-$2304 & 0.008 & \sci{7.9}{8\sx} & Curv  & Yes & Yes &  0.93 & \sci{1.6}{-10} \\ 
  J0117$+$5914 & 0.101 & \sci{1.6}{12} & NRICS & Yes & Yes &  3.27 & \sci{1.4}{-10} \\ 
  J1959$+$2048 & 0.002 & \sci{3.3}{8\sx} & Curv  & Yes & Yes &  0.41 & \sci{1.4}{-10} \\ 
  J1801$-$2451 & 0.125 & \sci{8.1}{12} & Curv  & Yes & Yes &  3.63 & \sci{1.3}{-10} \\ 
  \tableline
%   \enddata
%\end{deluxetable}
  \end{tabular}
  \end{center}
\end{table*}

\section{Numerical model}
 
Using the full numerical system described in Section 4 of Paper II,
we have run several
simulations of these objects.  These results confirm our conclusions
about the different regimes of pair production discussed above,
with the numerically calculated luminosity remaining within
approximately a factor of 3 of the simple analytic model, with the
numerical model always lower, due to the effects of pair creation on
the tails of the distribution.  Over the 25 brightest pulsars, the
numerical results are on average of a factor of 2.1 lower than the
theoretical results, although the brightest pulsars show more
variance.

The calculated pulsar spectra matched expectations, although in
several cases a low-intensity high-energy tail due to NRICS was
observed, extending up to $\gamma_{\PFF}$ with a power law exponent of
roughly -2.  In general, this tail contains a negligible portion of the
energy, namely the energy predicted by equation (\ref{eq:E_nr}).

We also ran the numerical model using alternative heating models.  We
found that, since in the brightest objects curvature emission sets the
PFF, the precise temperature model mattered little for the observed
objects.  Using different
models of the stellar temperature only changed the expected luminosity
by on the order of 20\%.  However, for the lower-luminosity ICS-dominated
pulsars, the temperature is far more important, as the ICS process
relies on the thermal photon bath, potentially allowing future
$\gamma$-ray observations to discriminate between thermal models.

\section{Discussion}

In Table \ref{table:observed}, we compare the observed flux, the flux
by the model of \citet{zhang00}, the flux predicted by the
semi-numerical model of this paper, and the flux computed by running
the full numerical cascade model for each of the observed $\gamma$-ray
pulsars.  The flux predicted for both models
was derived from the total luminosity using equation
(\ref{eq:lum_to_flux}).

\begin{table*}
  \caption{Observed and predicted pulsar fluxes
     \label{table:observed}}
  \begin{center}
  \begin{tabular}{lccccccc} \tableline \tableline
  \mycolhead{} & \mycolhead{$B$} & \mycolhead{PFF} &
    \mycolhead{$\phi_{obs}\tablenotemark{a}$} & \mycolhead{$\phi_{zhang}\tablenotemark{b}$} &
    \mycolhead{$\phi_{pred}$} & \mycolhead{$\phi_{num}$} \\
    \mycolhead{Pulsar} & \mycolhead{(Gauss)} &
    \mycolhead{mech.} &
    \mycolhead{(erg cm$^{-2}$ s$^{-1}$)} &
    \mycolhead{(erg cm$^{-2}$ s$^{-1}$)} &
    \mycolhead{(erg cm$^{-2}$ s$^{-1}$)} &
    \mycolhead{(erg cm$^{-2}$ s$^{-1}$)}
  \\ \tableline
  J0534$+$2200 \rule[0mm]{0mm}{2.5ex} & \sci{7.6}{12} & Curv  & \sci{1.3}{-8\sx} & \sci{5.9}{-9\sx} & \sci{4.8}{-9\sx} & \sci{1.8}{-9\sx} \\ 
  J0835$-$4510 & \sci{6.8}{12} & Curv  & \sci{9.9}{-9\sx} & \sci{2.3}{-8\sx} & \sci{1.6}{-8\sx} & \sci{5.0}{-9\sx} \\ 
  J0633$+$1746 & \sci{3.3}{12} & NRICS & \sci{3.9}{-9\sx} & \sci{2.6}{-8\sx} & \sci{9.5}{-9\sx} & \sci{4.5}{-9\sx} \\ 
  J1709$-$4428 & \sci{6.3}{12} & Curv  & \sci{1.3}{-9\sx} & \sci{1.3}{-9\sx} & \sci{9.2}{-10} & \sci{3.0}{-10} \\ 
  J1513$-$5908 & \sci{3.1}{13} & RICS  & \sci{8.8}{-10} & \sci{5.3}{-10} & \sci{4.0}{-13} & \sci{2.0}{-13} \\ 
  J1952$+$3252 & \sci{9.7}{11} & NRICS & \sci{4.3}{-10} & \sci{5.1}{-10} & \sci{3.3}{-10} & \sci{1.8}{-10} \\ 
  J1057$-$5226 & \sci{2.2}{12} & NRICS & \sci{2.9}{-10} & \sci{2.5}{-10} & \sci{7.6}{-11} & \sci{5.1}{-11} \\ 
  J1048$-$5832 & \sci{7.0}{12} & Curv  & \sci{2.5}{-10} & \sci{4.2}{-10} & \sci{2.9}{-10} & \sci{7.5}{-11} \\ 
%  \enddata
   \tableline
  \end{tabular}
  \tablenotetext{a}{\citet{thompson99}}
  \tablenotetext{b}{\citet{zhang00}}
  \end{center}
\end{table*}

The model of \citet{zhang00} assumed that $\alpha = 30$, decreasing
the expected flux, while invoking a proposed ICS instability to move the
acceleration region off the surface of the star.  These two effects
roughly cancel, leaving the
predictions of their model comparable to those of this model, which
effectively uses $\cos \alpha = 1$ and acceleration near the surface.
Due to the small numbers of reversed particles, we find no instability
in the acceleration zone, especially in the cases where, in the
language of this paper, curvature radiation sets the PFF.  Hence
we find no reason to raise the altitude of the acceleration zone.

The numerically calculated result is substantially smaller than either
the semi-numerical result or that of \citet{zhang00},
due to two major effects.  First, the numerically-calculated PFF is
lower than that predicted by the analytic model by approximately
25\% on average, which, since these pulsars operate in the low-level
quadratic portion of the accelerating potential, reduces the Lorentz
factor of the beam to approximately 60\% of its analytically
calculated value.  This occurs because photons on the exponential
tail of the curvature spectrum pair produce and create a sufficiently
dense plasma to short out the accelerating electric field at lower
altitudes than expected in cruder calculations \citep{arons79}.

The second reason is that some of the primary photon energy remains in
the generated pairs, rather than being re-radiated.  Both the analytic
model of this paper and that of \citet{zhang00} assume that {\em all} of the
energy emitted by the primary beam is eventually re-emitted in
$\gamma$-rays, due to the combination of synchrotron emission from
the created pairs and RICS extracting any remaining energy.  In the
numerical model, however, we calculate the pair spectrum itself and
can determine what fraction of the energy in the pairs is re-radiated
by RICS.

As a quick approximation to the effects of RICS on the pair spectrum, we 
assume that all particles with an energy loss length scale equal to or 
less than the stellar radius re-emit all of their energy as lower-energy 
$\gamma$-rays. The minimum Lorentz factor is the lowest at which thermal 
photons could be scattered into resonance with the field, $\gamma_{min} 
= \epsilon_B / \Delta \mu kT = 134.5 \, B_{12} T_6^{-1} \Delta \mu^{-1}$.
The maximum Lorentz factor is the point where the particle's energy 
loss scale is a stellar radius, $\gamma_{max} = \sci{4.05}{3} B_{12} 
T_6^{1/2}$.

With these corrections, we see that the predictions of the polar
cap model are low compared to the observations for most of the
observed $\gamma$-ray pulsars.  The difference is large
for the Crab, for which the prediction is low by a factor of 10,
and for the high-field object 1513-5908, which is low by three
orders of magnitude.  The other objects are typically low by a
factor of three.

The high-field object 1513-5908 clearly deserves further examination.
Not only is this magnetic field of this pulsar so large that the
applicability of this model is questionable, due to high-field
effects as discussed by \citet{harding97}, but preliminary studies
of the spatial variation of $\gamma$-ray emission across the polar 
cap suggest that the core of this pulsar's beam should be much brighter
than the edge field line considered in this model.

The neglect of spatial variation across the polar cap limits the
accuracy of these results; simple estimates suggest that including
those effects would increase the expected flux by 50\%, but more
detailed study is required to be more concrete.  Physically, the
general-relativistic acceleration is strongest at the center of the
pulsar, while the field line radius of curvature is smallest at the
edge.  Due to the smaller radius of curvature and gentler acceleration,
pair production due to ICS processes is far easier at the edges of
the polar cap than at the center, while pair production via curvature
radiation is more likely on the central field lines.  Together, these
effects combine to place the pair formation front at a higher altitude
in the center of the polar cap and lower near the edges, so that the 
central field lines are comparatively brighter than the edges.

These effects should help raise the predicted luminosities closer
to those observed, bringing them within the expected margins of
error for this study.  Further work on the variation of the
$\gamma$-ray emission across the polar cap is certainly required
before any firm conclusions can be drawn.

However, the polar cap model gives specific predictions for the
beaming which are not consistent with the observations.  First,
the $\gamma$-ray peaks should be in phase with the radio emission,
which is not typically the case.  Second, for all
of the observed pulsars, the curvature emission energy loss, 
equation (\ref{eq:gamma_s}), predicts that 40--60\% of the emitted
energy for these pulsars should be emitted within the first 10 km
above the surface.  This energy would go into a cone of angular width
$\theta_{emit} = 2 (3/2) 2^{1/2} \theta_c$ or
$\theta_{emit} = 3.5 \, P^{-1/2}$ degrees.  This is far more focused
than the observed pulsars, which have double beam profiles spread over
rotation phase 0.4, typically.  If the observed pulsars were all oriented
with the magnetic
axis, rotation axis, and line of sight all roughly parallel, such broad
profiles could be generated by a polar cap model.  Specific modeling of the
polar cap model's beaming properties require the angle between the
rotation axis and the magnetic axis to be less than 45 degrees and the
intrinsic opening angle of the beam to be as large as 30 degrees
\citep{harding98b}.  The restriction on obliquity is {\em a priori}
improbable, and is known to cause difficulty with the population
statistics of pulsars \citep{romani96}, while the large opening angle for
the intrinsic emission beam requires adding new features to the basic
dynamics of the polar cap model not supported by our calculations.

Given these beaming problems with the polar cap model, the
outer gap models of \citet{romani95,yadrioglu95,zhang97} remain strong
contenders for the EGRET pulsars.  These models easily produce beam
profiles which resemble the data, but are energetically more difficult.
For example, recent work by \citet{hirotani01} addresses the
over-abundance in TeV emission, but the emissivity in the GeV range
remains an open question.

However, even if the observed $\gamma$-ray emission originates in the
outer magnetosphere, the flux predicted to arise from the polar cap
itself is
not so far less than the observed fluxes as to be unobservable in the near
future.  With the advent of the new $\gamma$-ray instruments, the threshold
of sensitivity should be low enough that the polar cap $\gamma$-rays
can be observed.  Detection of these gamma rays, seen when looking down
the barrel of the radio gun, and comparison of their emission phase and 
profile with the radio counterpart, would be a firm test of the
long established, but not directly tested, theory of polar cap pair
creation which underlies 30 years of theorizing about the origin of pulsar
radio emission.

\section{Conclusion}

The primary conclusion of this paper is that the currently observed
$\gamma$-rays pulsars are {\em not} representative of the bulk of
pulsars.  Currently, only the brightest $\gamma$-ray pulsars are
observed; these pulsars are selected to be those where curvature
radiation operates efficiently, which is not true of
pulsars in general.

Most pulsars have a beam energy controlled by nonresonant inverse
Compton scattering (NRICS), rather than
curvature emission, and will emit significantly fewer $\gamma$-rays.
Locating examples of these objects will be a challenge to the new
generation of $\gamma$-ray telescopes.  The fainter the objects
which can be seen, the stronger the differences should be between the
NRICS cascades and the predictions of the curvature-based PFF models.

In these fainter objects, the separation of the mechanism producing
the PFF (and thus shorting out the accelerating potential) from the
mechanism producing most of the observable $\gamma$-rays should
reveal itself through differing spectral indices.  Due to the effects
of inverse Compton scattering, there should exist
two classes of observed $\gamma$-ray emission, one a raw curvature
spectrum characterized by a spectral index of $-2/3$, and the other a
saturated synchrotron response with a spectral index of $-3/2$.

For the currently observed $\gamma$-ray pulsars, the predicted fluxes 
are consistently low, although, due to the neglect of spatial variation 
across the polar cap, they are within the accuracy of this study. The 
beaming issues with the polar-cap model remain; for this reason, the 
outer-gap models are still strong contenders for the observed pulsars, 
despite the energetic problems of those models. Even if the currently 
observed emission is from an outer gap, improved $\gamma$-ray 
sensitivity should reveal the signal of the polar caps, which is at 
worst only a factor of 5 below the observed fluxes.

Further observations should also reveal the differences in the
luminosity classes of polar cap $\gamma$-rays, according to which
mechanism which creates the PFF, and the differences between the
two spectral regimes, providing a straightforward test of this
model.

\acknowledgements

The author would like to thank the referee for several helpful
suggestions.

\end{document}